\documentclass[a4paper,12pt,tightenlines,notitlepage,secnumarabic]{revtex4}
\usepackage{amsmath,amssymb,setspace}
\usepackage{graphicx}
\usepackage{geometry,layout}
\usepackage{newtxtext}
\usepackage{newtxmath}
\usepackage[normalem]{ulem}
\usepackage[utf8x]{inputenc}
\usepackage[nodayofweek,ddmmyy]{datetime}
\usepackage{color}
\definecolor{darkred}{rgb}{0.7,0,0}
\definecolor{darkgreen}{rgb}{0,0.3,0}
\definecolor{darkblue}{rgb}{0,0,0.7}
\definecolor{darkcyan}{rgb}{0,0.3,0.3}
\definecolor{darkmagenta}{rgb}{0.5,0,0.5}
\definecolor{lightgrey}{rgb}{0.7,0.7,0.7}
\definecolor{lightcyan}{rgb}{0,0.9,0.9}
\definecolor{lightblue}{rgb}{0.7,0.7,1.0}
\definecolor{lightyellow}{rgb}{1.0,1.0,0.75}
\definecolor{sepia}{rgb}{0.4,0.15,0.15}
\definecolor{palecyan}{rgb}{0.75,1,1}
\definecolor{palemagenta}{rgb}{1,0.75,1}

\usepackage[unicode,colorlinks,citecolor=darkblue,linkcolor=darkblue,bookmarks=true]{hyperref}

\newcommand{\eg}{{\it e.g.\,}}

\renewcommand{\le}{\leqslant}

\newcommand{\partd}[2]{\dfrac{\partial#1}{\partial#2}}

\newcommand{\ket}[1]{|#1\rangle}

\newcommand{\mean}[1]{\langle#1\rangle}

\graphicspath{{img/}}

\begin{document}
\title{Overcoming inefficient detection in sub-shot-noise absorption measurement and imaging}
\date{\today}

\newcommand{\prev}{\color{darkcyan}}
\newcommand{\ch}{\cosh}
\newcommand{\sh}{\sinh}

\author{E.~Knyazev}
\affiliation{LIGO, Massachusetts Institute of Technology, Cambridge, MA 02139, USA}

\author{F.~Ya.~Khalili}
\affiliation{Faculty of Physics, Lomonosov Moscow State University,  Moscow 119991, Russia,}
\affiliation{Russian Quantum Center, Skolkovo 143025, Russia}

\author{M.~V.~Chekhova}
\affiliation{Max-Planck-Institute for the Science of Light, Staudtstrasse 2, 91058 Erlangen, Germany}
\affiliation{Friedrich-Alexander-Universit\"at Erlangen-N\"urnberg, Staudtstrasse 7/B2, 91058 Erlangen, Germany,}
\affiliation{Faculty of Physics, Lomonosov Moscow State University,  Moscow 119991, Russia}

\begin{abstract}
Photon-number squeezing and correlations enable measurement of absorption with an accuracy exceeding that of the shot-noise limit. However, sub-shot noise imaging and sensing based on these methods require high detection efficiency, which can be a serious obstacle if measurements are carried out in ``difficult'' spectral ranges. We show that this problem can be overcome through the phase-sensitive amplification before detection. Here we propose an experimental scheme of sub-shot-noise imaging with tolerance to detection losses.
\end{abstract}

\maketitle

\section{Introduction}

Quantum metrology, one of the rapidly developing quantum technologies, uses quantum resources for overcoming the limits set by classical measurement methods~\cite{96a1BrKh, Giovannetti_Science_306_1330_2004}. Phase sensitivity is one example~\cite{Demkowicz_PIO_60-345_2015}, measurement of absorption~\cite{Moreau2017} is another one. Together with the spatial resolution, these two types of measurement form the basis for quantum imaging~\cite{Kolobov2007}. With spectral resolution, quantum-enhanced measurements of both types can be used for spectroscopy.  

\begin{figure}[]
  \centering
  \includegraphics[width=0.6\textwidth]{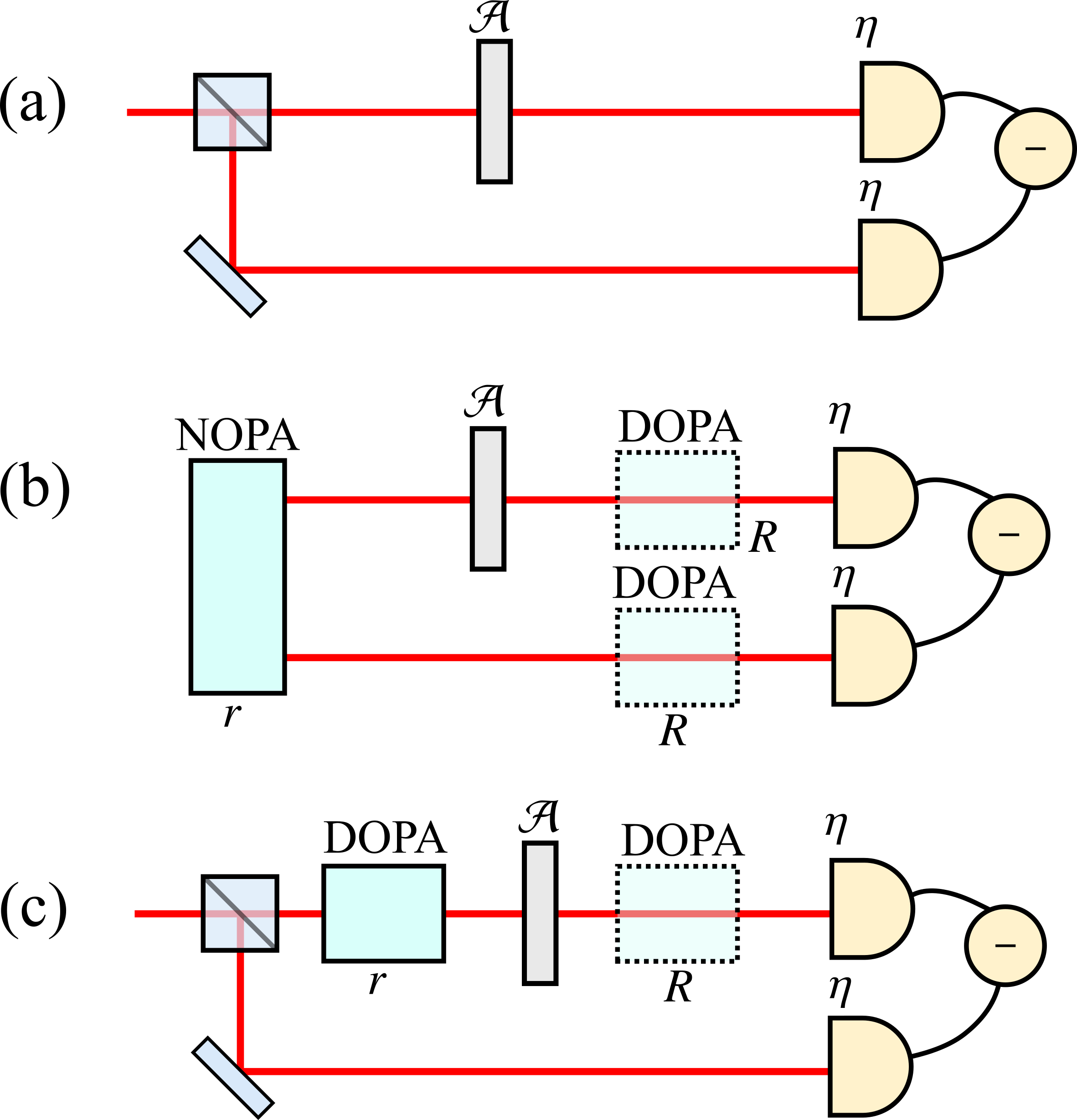}
  \caption{Schemes of quantum imaging and sensing. (a) Classical differential scheme. The object with absorption $\cal{A}$ is placed into one of the outputs of a 50\% beamsplitter, and the absorption is retrieved through intensity subtraction measurement. (b) Conventional scheme of sub-shot-noise quantum imaging. The object  is placed into one of the twin beams, emerging from a NOPA with the parametric gain $r$. Two photodetectors with equal quantum efficiencies $\eta$ measure the numbers of photons in signal and idler beams, and the difference is calculated.  (c) An alternative scheme, where only one beam is squeezed by a DOPA before probing the object, the other (much stronger) one being coherent. In schemes (b) and (c), additional DOPAs with gain $R$ (shown by dashed lines) can be placed into both arms (b) or into the signal arm (c) to overcome the detrimental effect of the detection loss.}
  \label{fig:imaging}
\end{figure}

In this work we will focus on the measurement of absorption, especially for the case of weakly absorbing objects. In classical optics, in order to measure a weak absorption $\mathcal{A}\ll1$, the object under test is placed into one of the output beams of a 50\% beamsplitter (Fig.\,\ref{fig:imaging}a). The second beam is used as a reference, for suppressing the effect of the amplitude fluctuations present in the incident beam. The absorption of the object is found by subtracting the output signals of the detectors placed into both channels. 

On the fundamental level, the sensitivity of absorption measurements is limited by the quantum fluctuations of the light intensity. In the simplest case of a coherent quantum state at the input, the sensitivity scales as $\Delta\mathcal{A}\sim1/\sqrt{N}$, see Eq.\,(\ref{DA_CR_coh_0}), where $N$ is the number of photons used~\cite{Jakeman1986,Monras2007}. This characteristic dependence, which originates from the Poissonian distribution of the photon number in the coherent state, is known as the {\em shot noise limit} (SNL). Although it can be improved by simply increasing the number of photons, this may not be an option in environmental or biological measurements where the use of high intensity should be avoided. 

More advanced quantum states of light could provide better sensitivity for weak illumination. The ideal case evidently corresponds to the Fock quantum states $\ket{N}$. In this case, $\Delta\mathcal{A}$ is considerably reduced as it gets multiplied by  a factor $\sqrt{\mathcal{A}}$, see Eq.\,(\ref{DA_CR_N_0}), ~\cite{Adesso_PRA_79_040305_2009}. The practical implementation of this idea uses twin beams produced by a non-degenerate  parametric amplifier (NOPA)  and relies on their high degree of photon-number correlation~\cite{Jedrkiewicz2004,  Brida2009, Agafonov2010}. In this scheme (Fig.\,\ref{fig:imaging}b), the absorbing object is placed into one of the beams (the signal one), with the second one serving as a reference. The  photon-number measurement in the reference beam projects the signal one into the quantum state with a well-defined $N$ (in the ideal case --- into the Fock state $\ket{N}$), enabling thus the sub-shot-noise sensitivity \cite{Jakeman1986,Tapster_PRA_44_3266_1991,Ribeiro1997,Whittaker_2017, Moreau2017}.  

A similar scheme was proposed in Ref.~\cite{Brambilla2008} for improving the signal-to-noise ratio (SNR) in imaging. In this scheme, the object to be imaged is placed into one of \textit{multimode} twin beams. The image is retrieved by subtracting the outputs of spatially resolving detectors placed into both beams. The SNR is then improved compared to the case of classical differential imaging due to the noise suppression below the SNL. This technique of sub-shot-noise imaging was demonstrated experimentally in works \cite{Brida2010,Samantaray2017}.

Instead of twin beams one can use a single sub-Poissonian squeezed beam, produced in a degenerate optical parametric amplifier (DOPA), as a probe \cite{Xiao_OL_13_476_1988}. A coherent beam fed from the same laser source can be used in this case as a reference (Fig.~\ref{fig:imaging}c). Although not suitable for imaging in the case of a single-mode beam, this scheme is convenient for spectroscopy. It was indeed used to enhance the sensitivity of spectroscopic measurements, with squeezed light from a parametric amplifier~\cite{Polzik1992} and from an amplitude-squeezed semiconductor laser~\cite{Kasapi2000}. This method can be very useful for sensing through absorption measurement, for instance, for monitoring the concentration of various gases~\cite{Hodgkinson2013}. 

For both these versions of sub-shot-noise quantum sensing, the increase in SNR strongly depends on the detection efficiency of the optical setup. In the experiments~\cite{Brida2010, Samantaray2017, Moreau2017, Whittaker_2017} performed in the visible range, the total detection efficiency of the setup exceeded $90\%$. However, in some cases the absorption should be measured in other spectral ranges, where detection is inefficient. Moreover, as we show further, in order to reveal the full potential of quantum sensing, the detectors inefficiency should not exceed the measured absorption. In the case of weak absorption measurements, $\mathcal{A}\lesssim10^{-2}$, this requirement can not be fulfilled with the existing photodetectors.

Here we show that sub-shot-noise quantum imaging can work in the presence of any loss provided that the beams carrying the information about the absorption are amplified before the detection using phase-sensitive parametric amplifiers (DOPAs), shown by dashed lines in Fig.~\ref{fig:imaging}b,c. This approach is similar to the strategy of overcoming loss in high-precision interferometry, first proposed in Ref.~\cite{Caves1981}, elaborated theoretically in Refs.~\cite{Yurke_PRA_A_33_4033_1986, Marino_PRA_86_023844_2012, Sparaciari_2016, 17a1MaKhCh}, and demonstrated experimentally in Refs.~\cite{Jing_APL_99_011110_2011, Kong_APL_102_011130_2013, Hudelist_NComms_5_3049_2014, 17a1MaLoKhCh}. It was also proposed as a method of ``quantum phase magnification'' \cite{Hosten2017}, and for ``enhanced quantum tomography''~\cite{Leonhardt_PRA_48_4598_1993, 17a1KnSpChKh}. 

In all these cases, the quadrature containing the signal is amplified and this way protected against loss. In the twin beam case which we explore here, the amplification is also phase-sensitive, but it does not matter which phase is amplified as long as it is the same for both twin beams. It is important that an ideal phase-sensitive amplifier does not introduce additional noise, and for this reason, it can be used for the noiseless amplification of images~\cite{Choi_PRL_83_1938_1999}. 

The present work considers the case where the image to be amplified is a sub-shot-noise one. In order to emphasize the effect, in our numerical estimates we will target very weak absorption $\mathcal{A}\approx10^{-5}$, typical, in  particular, for gas absorption measurements.

The paper is organized as follows. Section~\ref{sec:general} considers the fundamental bounds for the absorption measurement. In Section \ref{sec:Twin} we discuss sub-shot-noise quantum imaging using twin beams and the tolerance to inefficient detection provided by phase-sensitive amplification (Fig.~\ref{fig:imaging}b). Section \ref{sec:Squeezed} describes an alternative method, using a single squeezed coherent beam, also followed by phase-sensitive amplification for overcoming the detection loss (Fig.~\ref{fig:imaging}c). In Section \ref{sec:Exp} we outline possible experiments, and Section \ref{sec:Concl} is the conclusion.

\section{Cramer-Rao bounds}\label{sec:general} 

Statistics of the number of photons $n$ after an absorbing object are described by a conditional probability distribution $W(n/\mathcal{A})$ parameterized by the power absorption factor $\mathcal{A}$. The ultimate accuracy of estimating this parameter of the  probability distribution is given by the Cramer-Rao bound \cite{Kay2010},
\begin{equation}\label{CR_gen_org} 
  \Delta\mathcal{A}_{\rm CR} = \left\{
      \sum_{n=0}^\infty
        \frac{1}{W(n/\mathcal{A})}\left[\partd{W(n/\mathcal{A})}{\mathcal{A}}\right]^2
    \right\}^{-1/2} .
\end{equation} 
This equation describes the ideal case of a precise measurement of $n$ (that is,  photon counting with 100\% efficiency) as well as an ideal (lossless) preparation of the incident quantum state of light. However, it is easy to modify it to take into account the non-ideal detection efficiency and the non-ideal preparation. We model them by imaginary gray filters with the power transmissivities $\eta_d$ and $\eta_p$, located, respectively, after and before the object. In this case, the object transmissivity 
\begin{equation}
  \mathcal{T} = 1-\mathcal{A}
\end{equation} 
in Eq.\,\eqref{CR_gen_org} has to be replaced by the corresponding combined one, 
\begin{equation}\label{A_eta} 
  1-\mathcal{A}_\eta = \eta\mathcal{T} \,,
\end{equation} 
where
\begin{equation}
  \eta = \eta_p\eta_d
\end{equation} 
is the total quantum efficiency. This gives 
\begin{equation}\label{CR_loss_raw} 
  \Delta\mathcal{A}_{\rm CR} = \left\{
    \sum_{n=0}^\infty
      \frac{1}{W(n/\mathcal{A}_\eta)}
      \left[\partd{W(n/\mathcal{A}_\eta)}{\mathcal{A}}\right]^2
  \right\}^{-1/2} .
\end{equation} 
Combining Eqs.\,\eqref{CR_loss_raw} and \eqref{A_eta}, we obtain
\begin{equation}\label{CR_gen} 
  \Delta\mathcal{A}_{\rm CR} = \frac{1}{\eta}\left\{
      \sum_{n=0}^\infty
        \frac{1}{W(n/\mathcal{A}_\eta)}
        \left[\partd{W(n/\mathcal{A}_\eta)}{\mathcal{A}_\eta}\right]^2
    \right\}^{-1/2} .
\end{equation} 
Here we calculate this bound for two most important particular cases.

We start with a coherent state, which describes the laser radiation in the idealized case  where the excess (technical) noise is absent. This ``classical'' light is  characterized by the Poissonian photon-number distribution with some mean value $N_0$. After passing through the absorbing object and the imaginary filter, which models the detector inefficiency, the  photon-number distribution still remains Poissonian, but with a reduced mean photon number $(1-\mathcal{A}_\eta)N_0$:
\begin{equation}
  W(n/\mathcal{A}) = \frac{e^{-(1-\mathcal{A}_\eta)N_0}[(1-\mathcal{A}_\eta)N_0]^n}{n!}\,.
\end{equation} 
Substituting this distribution into Eq.\,\eqref{CR_gen}, we obtain the Cramer-Rao bound for coherent light: 
 
\begin{equation}\label{DA_CR_coh} 
  \Delta\mathcal{A}_{\rm CR\,coh} = \sqrt{\frac{\mathcal{T}}{\eta_dN}} \,,
\end{equation} 
where
\begin{equation}
  N = \eta_pN_0
\end{equation} 
is the mean photon number at the object. This limit is also known as the {\it shot noise} one. 

From a practical viewpoint, most important is the case of a highly transparent object and good preparation and detection efficiencies,
\begin{equation}\label{assump1} 
  \mathcal{A} \ll 1 \,, \quad  1-\eta_{p,d}\ll 1 \,.
\end{equation} 
In this case, the sensitivity is virtually not affected by the object absorption and by the preparation inefficiency,
\begin{equation}\label{DA_CR_coh_0} 
  \Delta\mathcal{A}_{\rm CR\,coh} = \frac{1}{\sqrt{N}} \,.
\end{equation} 

Consider now a photon-number (Fock) state $|N_0\rangle$. Taking into account that in the absorption measurements the information is encoded into the probe light intensity, this state should provide the best possible sensitivity. Indeed, for this state, the probability distribution for the number of detected photons is the binomial one,  
\begin{equation}\label{binom} 
  W(n/\mathcal{A}) = \frac{N_0!}{n!(N_0-n)!}\mathcal{A}_\eta^{N_0-n}(1-\mathcal{A}_\eta)^n
  \,,
\end{equation}
and the corresponding Cramer-Rao bound is 
\begin{equation}\label{DA_CR_N}
  \Delta\mathcal{A}_{\rm CR\,Fock} 
  = \sqrt{\frac{[\mathcal{A} + (1-\eta)\mathcal{T}]\mathcal{T}}{\eta_dN}}\,.
\end{equation} 
In the particular case of \eqref{assump1}, 
\begin{equation}\label{DA_CR_N_0}
  \Delta\mathcal{A}_{\rm CR\,Fock} 
  = \sqrt{\frac{\mathcal{A} + \epsilon^2}{N}}\,,
\end{equation} 
where 
\begin{equation}
  \epsilon^2 = \epsilon_p^2 + \epsilon_d^2
\end{equation} 
is the total inefficiency of the scheme and
\begin{equation}
  \epsilon_{p,d}^2 = \frac{1-\eta_{p,d}}{\eta_{p,d}} \,.
\end{equation} 
It is interesting that opposite to the phase measurements (see \eg \cite{17a1MaKhCh} and the references therein), Eqs.\,(\ref{DA_CR_N}, \ref{DA_CR_N_0}) still scale with the number of quanta as $1/\sqrt{N}$. However, the additional factor in the numerator could provide a significant sensitivity gain in the case of the weak ($\mathcal{A}\ll1$) absorption measurements with efficient preparation and detection ($1-\eta\ll1$). 

Given the same photon number $N$ of the beam hitting the object and the same values of $\eta_p$, $\eta_d$, the quantum advantage can be described  by the ratio $Q$ of the uncertainties for the shot-noise measurement $\Delta\mathcal{A}_{\rm CR coh}$ and the one using some nonclassical state of light,
\begin{equation}\label{QA} 
  Q = \frac{\Delta\mathcal{A}_{\rm CR\,coh}}{\Delta\mathcal{A}} \,.
\end{equation}
In the Fock state case ($\Delta\mathcal{A}=\Delta\mathcal{A}_{\rm CR\,Fock}$), it follows from Eqs.\,(\ref{DA_CR_coh_0}, \ref{DA_CR_N_0}) that this quantum advantage is 
\begin{equation}
  Q = \frac{1}{\sqrt{\mathcal{A} + \epsilon^2}} \,.
\end{equation} 
Clearly, the asymptotic value of this gain $1/\sqrt{\mathcal{A}}$ can be reached only if $\epsilon^2\ll\mathcal{A}$. In the opposite case, the gain is limited by the scheme inefficiency $\epsilon^2$.

\section{Twin-beams absorption measurement and its enhancement through phase-sensitive amplification} \label{sec:Twin}

Consider now the ``twin beams'' scheme shown in Fig.~\ref{fig:imaging}b. Let the twin beams be produced by a NOPA with the parametric gain $r$. The object, with the absorption $\mathcal{A}$, is placed into beam 1. Then both beams pass through DOPAs, which stretch some arbitrary but synchronized quadratures of these beams by $e^R$. After that, the beams are detected by two photodetectors.

There are various strategies of processing the output data. Here we consider two linear ones used, respectively, in \cite{Brida2010} and \cite{Moreau2017}. 

The first and simpler procedure is based on the estimator
\begin{equation}\label{est_simple} 
  \tilde{\mathcal{A}} = \frac{n_{d1} - n_{d2}}{G} \,,
\end{equation} 
where $n_{d1}$ and $n_{d2}$ are the photon numbers registered by the first and the second photodetectors and 
\begin{equation}
  G = \partd{\mean{n_{d1}}}{\mathcal{A}}
\end{equation} 
is the transfer function. The measurement error for this procedure is calculated in Appendix \ref{app:twin_beams}, see Eq.\,\eqref{delta_A_pdc_simple}. This general equation is quite cumbersome. In order to reveal its structure, we consider here the case \eqref{assump1} of a small absorption and good efficiency of the setup, assuming in addition for simplicity that the photon number is large,
\begin{equation}\label{big_N} 
  N\gg 1 \,.
\end{equation}
Also for simplicity, we present here the equations for only two most interesting cases: either the output DOPAs are not used at all ($R=0$), or they provide strong amplification,
\begin{equation}\label{big_R} 
  e^{2R} \gg 1 \,.
\end{equation} 

In the case of no amplification, $R=0$, Eq.\,\eqref{delta_A_pdc_simple} simplifies to 
\begin{subequations}\label{Delta2Apdc1}
  \begin{equation}\label{Delta2Apdc1_0}
    (\Delta\mathcal{A})^2 
    = \mathcal{A}^2 + \frac{\mathcal{A} + 2\epsilon^2}{N} \,.
  \end{equation}
  This limit differs from the Cramer-Rao bound \eqref{DA_CR_N_0} in two aspects. The first one is the additional term $\mathcal{A}^2$, which stems from the asymmetry introduced into the twin beams by the absorption and, as we will show below, can be removed by using an estimator more advanced than \eqref{est_simple}. Second, the terms imposed by the preparation and detector inefficiency are twice as large due to the two beams used in this protocol.

  In the case \eqref{big_R} of strong amplification, 
  \begin{equation}\label{Delta2Apdc1_R} 
    (\Delta\mathcal{A})^2 = 2\left[
        \mathcal{A}^2 + \frac{\mathcal{A} + 2(\epsilon_p^2 + \epsilon_d^2e^{-2R})}{N}
      \right]
      + \frac{1}{N^2} \,.
  \end{equation} 
\end{subequations}
One can see that in this case the detection inefficiency is suppressed by the factor $e^{-2R}$. At the same time, the additional photon-number fluctuations originating from the imbalance due to object are enhanced by the DOPAs, thus adding: (i) yet another factor 2, and (ii) the additional term $1/N^2$. 

In Ref.~\cite{Moreau2017}, a more sophisticated but still linear estimator was proposed:
\begin{equation}\label{est_smart} 
  \tilde{\mathcal{A}} = \frac{n_{d1} - kn_{d2}}{G} \,,
\end{equation} 
where the factor $k$ should be optimized to provide the minimum of $\Delta\mathcal{A}$. The corresponding minimized $\Delta\mathcal{A}$ is also calculated in Appendix \ref{app:twin_beams}, see Eq.\,\eqref{delta_A_pdc_smart}. The resulting asymptotic equations for the absorption uncertainty and the quantum advantage for the same ``no amplification'' and ``strong amplification'' cases as above are, respectively, 
\begin{subequations}\label{Delta2Apdc2} 
  \begin{equation}\label{Delta2Apdc2_0} 
    (\Delta\mathcal{A})^2  = \frac{\mathcal{A} + 2\epsilon^2}{N} \,, \qquad 
    Q = \frac{1}{\sqrt{\mathcal{A} + 2\epsilon^2}} \,,
  \end{equation} 
  and 
  \begin{equation}\label{Delta2Apdc2_R}
    (\Delta\mathcal{A})^2 
    = 2\left[\frac{\mathcal{A} + 2(\epsilon_p^2 + \epsilon_d^2e^{-2R})}{N}\right]
      + \frac{1}{N^2} \,,
    \qquad 
    Q = \frac{1}{\sqrt{2[\mathcal{A} + 2(\epsilon_p^2 + 2\epsilon_d^2e^{-2R}] + 1/N}} \,.
  \end{equation} 
\end{subequations}
Equations \eqref{Delta2Apdc2} differ from the previous ones \eqref{Delta2Apdc1} by the absence of the additional term $\mathcal{A}^2$. The price for this is that this procedure requires a priori information on $\mathcal{A}$.

\begin{figure}
  \centering
  \includegraphics[width=0.65\textwidth]{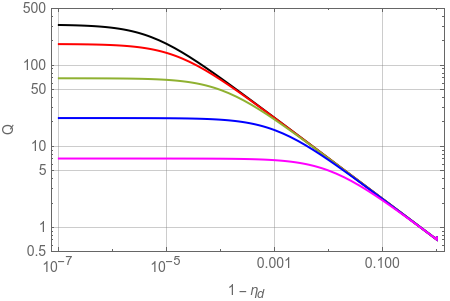}
  \caption{The quantum advantage $Q$ of the `twin beams' scheme, see Eqs.\,(\ref{QA}, \ref{delta_A_pdc_smart}), for the case of no parametric amplification ($R=0$) as a function of detection efficiency $\eta_d$ for $\epsilon_p^2=0$ (black), $10^{-5}$ (red), $10^{-4}$ (green), $10^{-3}$ (blue), and $10^{-2}$ (magenta). The object absorption is $\mathcal{A}=10^{-5}$ and the mean photon number is $N=10^7$.}
  \label{fig:Q0PDC}
\end{figure}

\begin{figure}[]
  \centering
  \includegraphics[width=0.65\textwidth]{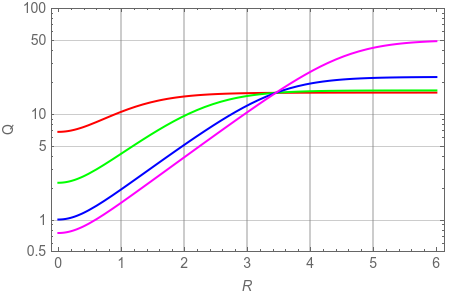}
  \caption{The quantum advantage $Q$ of the `twin beams' scheme in the presence of additional parametric amplification, see Eqs.\,(\ref{QA}, \ref{delta_A_pdc_smart}), as a function of the gain $R$ for different values of the detection efficiency: $\eta=0.99$ (red), $0.9$ (green), $0.5$ (blue), and $0.1$ (magenta). The mean photon number is $N=10^7$ and the absorption is $\mathcal{A}=10^{-5}$.}
  \label{fig:QPDC}
\end{figure}

The quantum advantage $Q$ achieved using the estimator \eqref{est_smart} in the case of $R=0$ is shown in Fig.\,\ref{fig:Q0PDC} as a function of the detection inefficiency $\epsilon_d^2$ for various values of the preparation inefficiency [note that the plots in Fig.\,\ref{fig:Q0PDC}, as well as in Fig.\,\ref{fig:QPDC} below are drawn using the exact equations \eqref{delta_A_pdc_smart} and \eqref{delta_A_coh_sqz}, respectively]. Among them, $\epsilon_p^2=10^{-3}$ can be considered as a realistic one. It corresponds to the absorption inside the nonlinear crystal on the order of $10^{-4}\rm{mm}^{-1}$~\cite{Dmitriev1999} and reflection at each surface of the crystal about $10^{-4}$~\cite{Newlight}. Clearly, $Q$ is above unity only for sufficiently high detection efficiency. Similarly to the Fock state case (see Sec.\,\ref{sec:general}), the quantum advantage reaches its maximal value $1/\sqrt{\mathcal{A}}$ only if $\epsilon_{d,p}^2\ll\mathcal{A}$. In the opposite case, the quantum advantage is limited by the scheme inefficiency:
\begin{equation}
  Q \le \frac{1}{\sqrt{2}\,\epsilon} \,. 
\end{equation} 

However, as one can see from Eqs.\,(\ref{Delta2Apdc1_R}, \ref{Delta2Apdc2_R}), 
by increasing the amplification $R$ the effect of the {inefficient detection can be made arbitrarily small. In the practical case of $\mathcal{A}\ll\epsilon^2$, the corresponding sensitivity gain can more than compensate for the above-mentioned  additional factor 2 before $\mathcal{A}$ and the additional term $1/N^2$ in the equation for $\Delta\mathcal{A}$. This is demonstrated by Fig.\,\ref{fig:QPDC}, where the quantum advantage achieved with the estimator \eqref{est_smart} is plotted as a function of the parametric gain $R$ for several values of the detectors efficiency, from the very optimistic to the very low ones. One can see that for any quantum efficiency the asymptotic value of the quantum advantage, 
\begin{equation}
  Q = \frac{1}{\sqrt{2(\mathcal{A}+2\epsilon_p^2)+1/N}},
\end{equation} 
can be reached, provided the sufficiently strong parametric amplification. 

It is interesting to note that in the case of a low quantum efficiency, this value can be even exceeded, see the blue and magenta curves in Fig.~\ref{fig:QPDC}. This is because the baseline value $\Delta\mathcal{A}_{\rm CR\,coh}$ is also affected by the detectors inefficiency, and this effect becomes significant if $\epsilon^2_d\gtrsim1$, see Eq.\,\eqref{DA_CR_coh}.

\section{Sub-shot-noise sensing with squeezed  light} \label{sec:Squeezed}

An alternative way to overcome the shot noise limit is to use another quantum state with reduced photon-number fluctuations, namely the squeezed coherent state. Consider the scheme shown in Fig.~\ref{fig:imaging}c. Here a coherent beam is split on a beamsplitter, so that part (top) is amplitude squeezed in a degenerate parametric amplifier (DOPA) with the parametric gain $r$ and used as a probe beam with the photon number $N$, and another one (bottom) is used as a reference. Then the probe beam passes through another DOPA, which has the parametric gain $R$.

Note that while the optical power in the probe beam could be limited by  the probed object fragility, there is no such limitation for the reference beam. Therefore, it is reasonable to have the reference beam much stronger than the signal one, in order to suppress the reference beam shot noise. In this case, an asymmetric beamsplitter has to be used, and the reference photodetector output has to be proportionally scaled down. We assume this case here, taking into account the quantum noise of the signal beam only. 

The measurement uncertainty for this scheme is calculated in Appendix \ref{app:coh_sqz}, see Eq.\,\eqref{delta_A_coh_sqz}. Similarly to the previous section, we consider several characteristic asymptotic cases. 

We start with the baseline case of $R=0$. If the input squeezing is absent as well, then Eq.\,\eqref{delta_A_coh_sqz} reduces to the shot-noise limit \eqref{DA_CR_coh}. Suppose that the input optical field is strongly squeezed,
\begin{equation}\label{big_r} 
  e^{-2r} \gg 1, 
\end{equation} 
and assume also conditions (\ref{assump1}, \ref{big_N}). In this case it follows from Eq.\,\eqref{delta_A_coh_sqz} that  
\begin{subequations}
  \begin{equation}
    (\Delta\mathcal{A})^2 
    = \frac{e^{-2|r|} + \mathcal{A} + \epsilon^2}{N} + \frac{e^{4|r|}}{8N^2} \,
  \end{equation}
  and
  \begin{equation}
    Q = \frac{1}{\sqrt{e^{-2|r|} + \mathcal{A} + \epsilon^2 + \dfrac{e^{4|r|}}{8N}}} \,.
  \end{equation} 
\end{subequations}
One can see that the optimal value of the input squeezing exists which provides the minimum of the initial uncertainty of the number of quanta and therefore the minimum of  $\Delta\mathcal{A}$. This optimal squeezing corresponds to 
\begin{equation}\label{opt_coh_0}
  e^{2|r|} = (4N)^{1/3},
\end{equation} 
which gives
\begin{subequations}\label{Delta2Acoh_0} 
  \begin{gather}
    (\Delta\mathcal{A})^2 = \frac{\mathcal{A} + \epsilon^2}{N} + \frac{3}{2^{5/3}N^{4/3}}
      \,, \label{DA_coh_0} \\
    Q = \frac{1}{\sqrt{\mathcal{A} + \epsilon^2 + \dfrac{3}{2^{5/3}N^{1/3}}}} \,.
  \end{gather} 
\end{subequations}
Note that the first two terms in Eq.\,\eqref{DA_coh_0} coincide with the Cramer-Rao bound for a Fock state \eqref{DA_CR_N_0}. The third term originates from the photon-number fluctuations in the incident light, which can not be made arbitrary small within the constraints of Gaussian (displacement and squeezing) transformations. Note that for the values of $\mathcal{A}$ and $N$ that we use in this paper, it is this term that dominates if the losses are small, $\epsilon_{p,d}^2\lesssim 10^{-3}$. At the same time, in the more realistic  case of stronger losses, the quantum advantage for the squeezed coherent state case can be $\sqrt{2}$ higher than in the twin-beam case, compare Eqs.\,\eqref{Delta2Apdc2_0} and \eqref{Delta2Acoh_0}. The reason for this is evident: the former scheme requires only one noisy channel while the latter, two of them.

\begin{figure}
  \centering
  \includegraphics[width=0.65\textwidth]{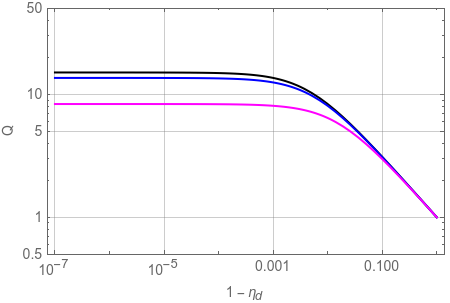}
  \caption{The quantum advantage $Q$ of the ``squeezed coherent'' scheme, see Eqs.\,(\ref{QA}, \ref{delta_A_coh_sqz}), for the case of no parametric amplification ($R=0$) as a function of detection efficiency $\eta_d$ for $\epsilon_p^2=0$ (black), $10^{-3}$ (blue), and $10^{-2}$ (magenta). The object absorption is $\mathcal{A}=10^{-5}$ and the mean photon number is $N=10^7$.}
  \label{fig:Q0coh}
\end{figure}

The quantum advantage for the squeezed coherent input state in the case of $R=0$ is shown in Fig.\,\ref{fig:Q0coh} as a function of the detection inefficiency $\epsilon_d^2$ for various values of preparation inefficiency, compare with the corresponding plots for the twin-beam scheme shown in Fig.\,\ref{fig:Q0PDC}. 

In the ``strong amplification'' case of \eqref{big_R}, assuming also the asymptotic conditions (\ref{assump1}, \ref{big_N}, \ref{big_r}), Eq.\,\eqref{delta_A_coh_sqz} can be reduced to 
\begin{equation}\label{D2A_coh_asy} 
  (\Delta\mathcal{A})^2 
  = \frac{e^{-2|r|} + \mathcal{A} + \epsilon_p^2 + \epsilon_d^2e^{-2R}}
      {N - \dfrac{e^{2|r|}}{4}} 
  + \frac{e^{4|r|-8R}}{8\left(N - \dfrac{e^{2|r|}}{4}\right)^2} \,.
\end{equation} 
This equation shows once again that the detector inefficiency can be overcome by the amplification.

Dependence of $\Delta\mathcal{A}$ on the input squeeze factor $r$ is non-trivial. However, in all reasonable practical cases, one can assume that 
\begin{equation}\label{mod_sqz}
  e^{2|r|}\ll 4N \,,
\end{equation} 
which gives
\begin{equation}
  (\Delta\mathcal{A})^2 
  = \frac{e^{-2|r|} + \mathcal{A} + \epsilon_p^2 + \epsilon_d^2e^{-2R}}{N} 
    + \frac{e^{4|r|-8R}}{8N^2} \,.  
\end{equation} 
The minimum of this equation in $|r|$ is provided by 
\begin{equation}\label{opt_coh_R_asy}
  e^{2|r|} = (4N)^{1/3}e^{8R/3} \,, 
\end{equation} 
which gives the following simple equations:
\begin{subequations}\label{Delta2Acoh_R}
  \begin{gather}
    (\Delta\mathcal{A})^2 
      = \frac{\mathcal{A} + \epsilon_p^2 + \epsilon_d^2e^{-2R}}{N} 
        + \frac{3e^{-8R/3}}{2^{5/3}N^{4/3}} \,, \label{D2Acoh_R}\\
    Q = \frac{1}{\sqrt{
            \mathcal{A} + \epsilon_p^2 + \epsilon_d^2e^{-2R} 
            + \dfrac{3e^{-8R/3}}{2^{5/3}N^{1/3}} 
          }} \,. \label{Qcoh_R}
  \end{gather} 
\end{subequations}
Then it follows from Eqs.\,(\ref{mod_sqz}, \ref{opt_coh_R_asy}) that in this case
\begin{equation}\label{small_R} 
  e^{4R} \ll 4N \,.
\end{equation} 
Comparison of Eqs.\eqref{Delta2Acoh_R} with \eqref{Delta2Acoh_0} shows that not only the detection inefficiency term, but also the one which stems from the Gaussianity of the input quantum state [the last one in \eqref{D2Acoh_R}] is suppressed by the amplification. 

In comparison with the twin-beam scheme of Sec.\,\ref{sec:Twin}, the effect of the object absorption on the quantum advantage is reduced by a factor of $\sqrt{2}$, and the effect of the  preparation and detection losses, by a factor of 2, compare Eqs.\,\eqref{Delta2Apdc2_R} and \eqref{Qcoh_R}. On the other hand, the last term in Eq.\,\eqref{Qcoh_R} is always larger than its counterpart in Eq.\,\eqref{Delta2Apdc2_R} due to the condition \eqref{small_R}. However, for the realistic parameters that we use in this paper, this term is smaller than the previous ones, making the ``squeezed coherent'' scheme more sensitive.

Figure~\ref{fig:SC_amp} shows the quantum advantage $Q$ and the corresponding numerically optimized  squeeze factor $r$  for the `squeezed coherent' scheme as a function of amplification $R$ for the same values of the detection efficiency and the other parameters as in Fig.\,\ref{fig:QPDC}. 
\begin{figure}[]
  \centering
  \includegraphics[width=0.65\textwidth]{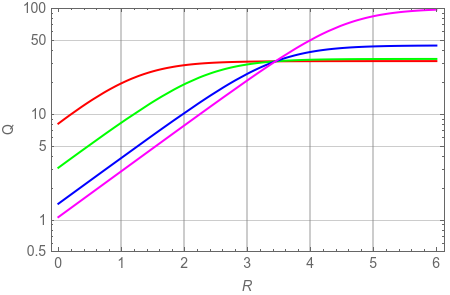}
  \includegraphics[width=0.65\textwidth]{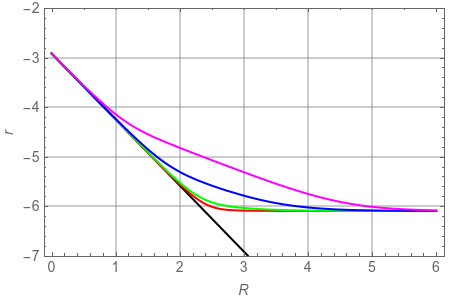}
  \caption{The quantum advantage $Q$ (top) and the corresponding numerically optimized squeeze factor $r$ (bottom) for the ``squeezed coherent'' scheme, see Eqs.\,(\ref{QA}, \ref{delta_A_coh_sqz}), as a function of the gain $R$ for different values of the detection efficiency: $\eta=0.99$ (red), $0.9$ (green), $0.5$ (blue), and $0.1$ (magenta). The straight black line in the bottom plot shows the asymptotic  \eqref{opt_coh_R_asy}. The mean photon number is $N=10^7$ and the absorption is $\mathcal{A}=10^{-5}$. }
  \label{fig:SC_amp}
\end{figure}

\section{Experimental implementation} \label{sec:Exp}

From the above analysis, it follows that the scheme with a squeezed coherent beam provides a better sensitivity in the absorption measurement than the twin-beams one. This scheme can be relatively easy implemented using a cavity-based DOPA, where the state-of-the-art squeezing reaches 15 dB~\cite{Vahlbruch_PRL_117_110801_2016}. The same device can serve for the amplification of the beam after the object to overcome the detection losses. This single-mode scheme will be suitable for sub-shot-noise spectroscopy or sensing without spatial resolution.

More challenging is to overcome detection loss in sub-shot-noise imaging, where spatially multimode beams are involved. A sub-Poissonian multimode beam can be produced from multimode twin beams through heralding~\cite{Iskhakov2016} but the resulting amount of photon-number squeezing is not high because of the inefficiency of the detector used in this method. Much more promising is to utilize the twin-beam squeezing available in spatially multimode traveling-wave NOPAs~\cite{Brida2010,Iskhakov2016,Samantaray2017} where the noise reduction down to $7.8$ dB has been reported~\cite{Iskhakov2016}.

The principal scheme of a possible experiment is shown in Fig.\,\ref{fig:ExpAlt}. Multimode twin beams are produced by a phase-insensitive traveling-wave NOPA. A certain technical difficulty arises from the necessity to provide phase-sensitive amplification with the same phase in each beam. To avoid matching the mode structure of phase sensitive amplifiers in both beams, one can take advantage of the transformation between phase sensitive and phase-insensitive amplification through mode conversion \cite{Choi_PRL_83_1938_1999}.
\begin{figure}[]
  \centering
  \includegraphics[width=0.99\textwidth]{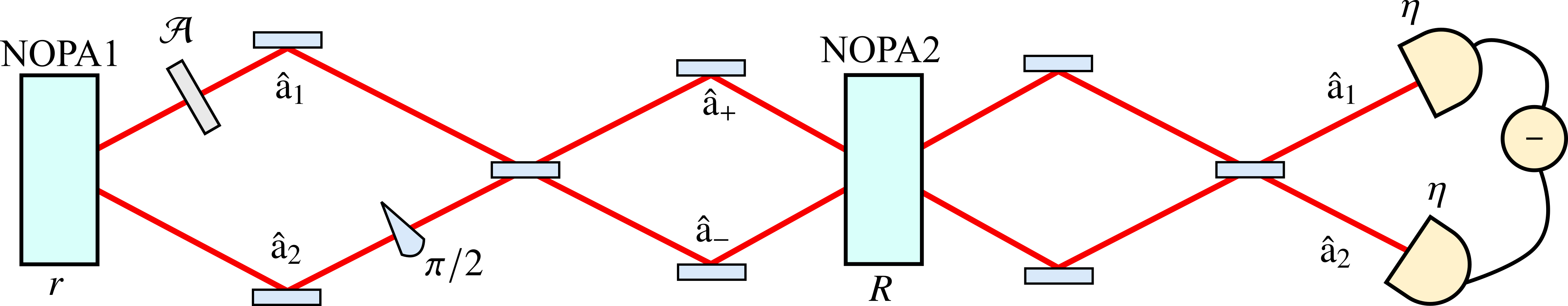}
  \caption{Scheme of a possible experiment. NOPA1 generates beams in modes $a_1$ and $a_2$, and a weakly absorbing object is placed into mode $a_1$. After the beam in mode $a_2$ acquires a $\pi/2$ phase shift, both beams are overlapped on a $50\%$ beamsplitter to form modes $a_\pm=(a_1\pm i a_2)/\sqrt{2}$ at the output. These two modes are amplified by NOPA2, which is equivalent to the phase-sensitive amplification of modes $a_1$ and $a_2$. In the end, photon numbers in modes $a_1$ and $a_2$ should be measured, which requires overlapping the modes on another $50\%$ beamsplitter.}
  \label{fig:ExpAlt}
\end{figure}

The operation of the NOPA producing the twin beams is described by the Hamiltonian
\begin{equation}
\hat{H}=i\hbar\gamma\hat{{\rm a}}_1^\dagger\hat{{\rm a}}_2^\dagger+h.c.,
\label{eq:HamPI}
\end{equation}
where $\gamma$ quantifies the interaction strength.

Phase sensitive amplification with the same phase and the same parametric gain in modes $\hat{{\rm a}}_1,\,\hat{{\rm a}}_2$ after the absorbing object is described by the Hamiltonian
\begin{equation}
\hat{H}=i\hbar\Gamma
  \left[(\hat{{\rm a}}_1^\dagger)^2 +(\hat{{\rm a}}_2^\dagger)^2\right]+h.c.,
\label{eq:HamPS}
\end{equation}
with the interaction strength characterized by $\Gamma$. This Hamiltonian can be represented as one of a phase-insensitive amplifier,
\begin{equation}
\hat{H}=i\hbar\Gamma\hat{{\rm a}}_+^\dagger\hat{{\rm a}}_-^\dagger + h.c.,
\label{eq:HamRL}
\end{equation}
for the modes 
\begin{equation}
  \hat{{\rm a}}_\pm = \frac{\hat{{\rm a}}_1 \pm i\hat{{\rm a}}_2}{\sqrt{2}} \,.
\end{equation}
The modes $\hat{{\rm a}}_{1,2}$ can be transformed into $\hat{{\rm a}}_\pm$ by using a 50\% beamsplitter, with the $\pi/2$ phase shift introduced into $\hat{{\rm a}}_2$. After the second NOPA, a second beamsplitter should be used to convert the ``$\pm$'' modes back into the ``$1,2$'' ones. 

\begin{figure}[]
  \centering
  \includegraphics[width=0.6\textwidth]{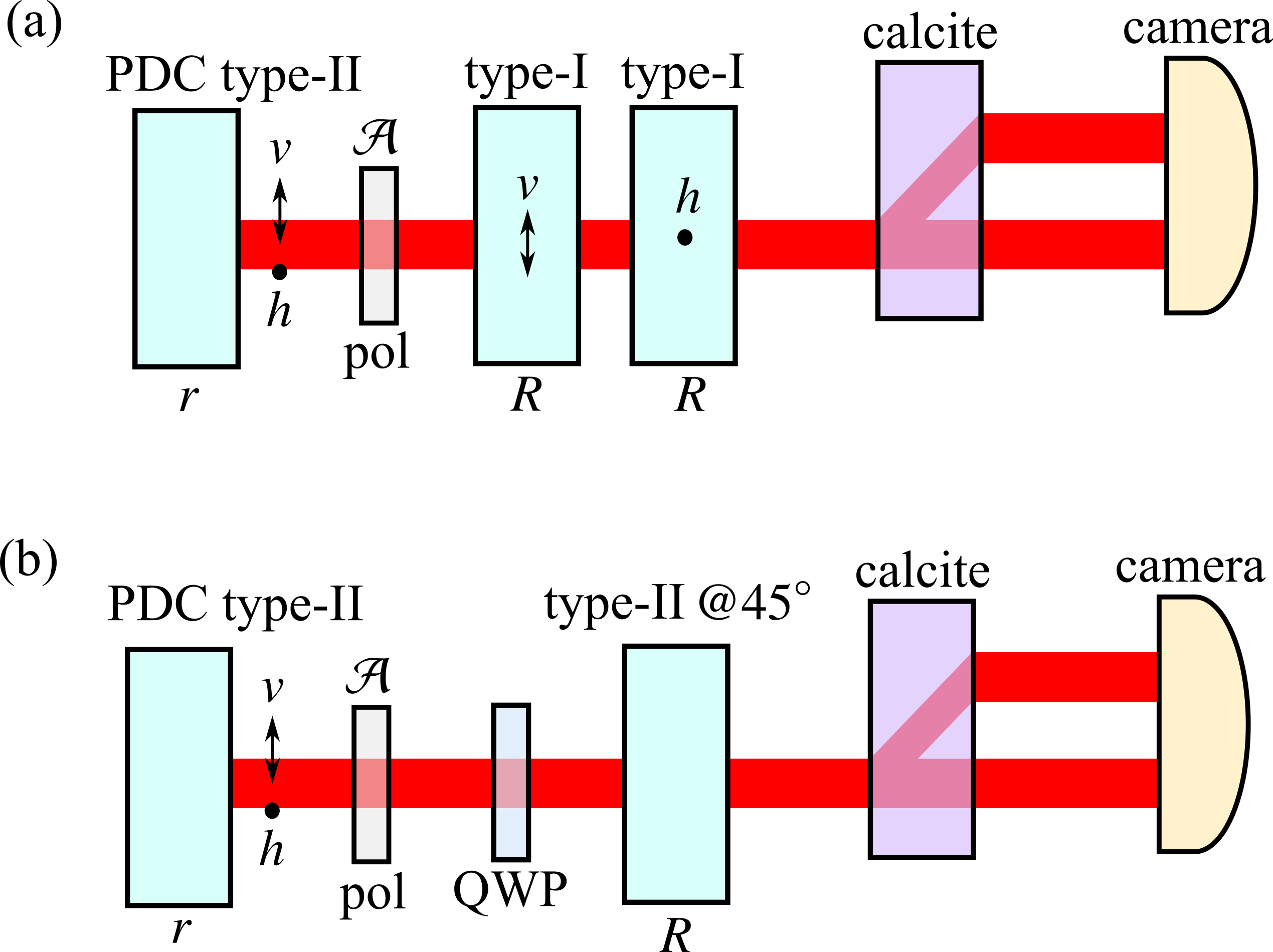}
  \caption{Polarization implementation of the experiment on sub-shot-noise imaging.}
  \label{fig:Exp}
\end{figure}

Although Fig.~\ref{fig:ExpAlt} looks very complicated, the same idea can be implemented much simpler using polarization optics (Fig.\,\ref{fig:Exp}). The vertically and horizontally polarized twin beams are produced through type-II parametric down-conversion, and the object is absorbing just one polarization. As phase sensitive amplifiers, one can use type-I parametric amplifiers placed into both beams. The first one is then acting just on the vertically polarized beam, mode v, and the second one, just on the horizontally polarized beam, mode h (panel a). In the end, both beams are detected separately, for instance, by different sensitive areas of the same camera after a birefringent beamsplitter. To retrieve the image, the intensity distributions obtained by the camera should be subtracted pixel-by-pixel.

However, it is experimentally challenging to provide amplification with the same phase for both polarizations. A solution is to use, as in Fig.~\ref{fig:ExpAlt}, instead of phase-sensitive amplification of vertical and horizontal polarizations (type-I OPA), a phase-insensitive two-mode amplification (type-II OPA), shown in panel b. This NOPA should amplify modes $(h\pm iv)/\sqrt{2}$, i.e., right- and left-circularly polarized modes. Because a type-II OPA can only amplify linearly polarized beams, it has to be preceded by a quarter wave plate placed into both beams. Note that the type-II OPA should have the polarization direction tilted by $45^\circ$.  

\section{Conclusion} \label{sec:Concl}

We have considered two methods of sub-shot-noise measurement of weak absorption, one using twin beams and the other one, squeezed coherent light. Both are ``substitutes'' for the ideal case where an absorptive object is probed by a Fock state, which has zero uncertainty in the photon number. The ``squeezed coherent'' scheme is easier to implement, with a single-mode coherent beam and a single-mode squeezer. However, while it is suitable for sensing or absorption measurement, for instance, in spectroscopy, it cannot be used for imaging, where the object should be illuminated by a spatially multimode beam. In this case, the `twin-beams' scheme is more convenient: a traveling-wave NOPA is always strongly multimode, unless special measures are taken. Both methods give advantage over the classical differential method, the ``squeezed coherent'' one providing a factor of $\sqrt{2}$ better performance than the other one and reaching the Cramer-Rao bound for Fock states in the ideal case of infinite squeezing and no loss. 

Meanwhile, both schemes turn out to be useless whenever the detection efficiency is low. This will be the case if imaging, sensing, or absorption spectroscopy is carried out in the mid-infrared or even terahertz spectral range. As we show, a way around it is to apply phase sensitive amplification before detection. We show that at any value of the detection efficiency, by sufficiently amplifying the beams after the absorbing object one can reach the same quantum advantage as in the case of high efficiency. 

The experimental implementation of this technique is most simple in sensing or spectroscopy, where no spatial resolution is required. Single-mode cavity-based OPAs can be used in this case, both for generating a squeezed coherent beam and for amplifying it after the object under study. For imaging, we consider a polarization setup with a traveling-wave type-II OPA producing orthogonally polarized twin beams. Another type-II OPA preceded by a quarter-wave plate can then be used for the amplification before detection. 

This work was supported by the joint DFG-RFBR project CH1591/2-1 - 16-52-12031 NNIOa. E.K. and F.K. acknowledge the financial support of the RFBR grant 16-52-10069.

\appendix

\section{Detection inefficiency}

We model the detectors inefficiency by imaginary beamsplitters with the power transmissivity $\eta_d$:
\begin{equation}
  \hat{{\rm d}}_{1,2} 
  = \sqrt{\eta_d}\,\hat{{\rm c}}_{1,2} + \sqrt{1-\eta_d}\,\hat{{\rm u}}_{1,2}
  = \sqrt{\eta_d}(\hat{{\rm c}}_{1,2} + \epsilon_d\hat{{\rm u}}_{1,2}) \,,
\end{equation} 
where $\hat{{\rm c}}_{1,2}$ are the fields at the input of the detectors, $\hat{{\rm d}}_{1,2}$ are the effective input fields of the corresponding imaginary lossless detectors, and $\hat{{\rm u}}_{1,2}$ are vacuum fields. 

It is easy to show that the statistics of the effective input fields depend on the statistics of the real ones as 
\begin{subequations}\label{stat_c2d} 
  \begin{gather}
    \mean{\hat{n}_{d1,2}} = \eta_d\mean{\hat{n}_{c1,2}} \,, \\
    \mean{(\delta\hat{n}_{d1,2})^2} 
      = \eta_d^2[\mean{(\delta\hat{n}_{c1,2})^2} + \epsilon_d^2\mean{\hat{n}_{c1,2}}] \,, \\
    \mean{\delta\hat{n}_{d1}\cdot\delta\hat{n}_{d2}}
      = \eta_d^2\mean{\delta\hat{n}_{c1}\cdot\delta\hat{n}_{c2}} \,,
  \end{gather}
\end{subequations}
where
\begin{subequations}
  \begin{gather}
    \hat{n}_{c1,2} = \hat{{\rm c}}_{1,2}^\dagger\hat{{\rm c}}_{1,2} \,, \\
    \hat{n}_{d1,2} = \hat{{\rm d}}_{1,2}^\dagger\hat{{\rm d}}_{1,2} \,, \\
    \delta\hat{n}_{c1,2} = \hat{n}_{c1,2} - \mean{\hat{n}_{c1,2}} \,, \\
    \delta\hat{n}_{d1,2} = \hat{n}_{d1,2} - \mean{\hat{n}_{d1,2}} \,.
  \end{gather}
\end{subequations}

\section{Twin beams}\label{app:twin_beams}

\begin{figure}
  \includegraphics[width=0.85\textwidth]{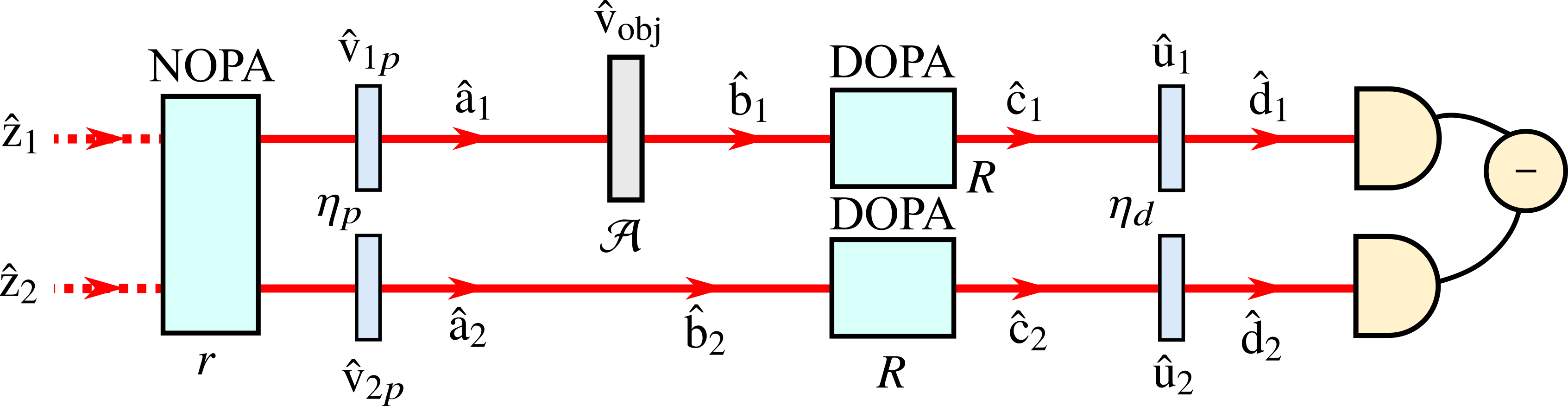}
  \caption{Measurement of absorption using twin-beams scheme with amplification before detection. The object under study is placed into the first beam, and both beams are simultaneously amplified before the direct detection. The state preparation loss is modeled by means of two identical beamsplitters with power transmissivity $\eta_p$, and the detection loss, by a similar beamsplitter $\eta_d$. The pump of the non-linear crystals is not depicted.}
  \label{fig:scheme_PDC_fields} 
\end{figure} 

The annihilation operators for two vacuum input fields (see Fig.\,\ref{fig:scheme_PDC_fields}) are $\hat{{\rm z}}_{1,2}$\,. The non-ideal NOPA transforms them into
\begin{equation}
  \hat{{\rm a}}_{1,2} 
  = \sqrt{\eta_p}(\hat{{\rm z}}_{1,2}\ch r + \hat{{\rm z}}_{2,1}^\dagger\sh r)
      + \sqrt{1-\eta_p}\,\hat{{\rm v}}_{1p,2p}\,,
\end{equation}
where $\hat{{\rm v}}_{1p}$, $\hat{{\rm v}}_{2p}$ are two independent vacuum fields. 

The object partly absorbs the first beam, leaving the second one unchanged:
\begin{subequations}
  \begin{gather}
    \hat{{\rm b}}_1 
      = \sqrt{\mathcal{T}}\hat{{\rm a}}_1 + \sqrt{\mathcal{A}}\hat{{\rm v}}_{\rm obj}
      = \sqrt{\mathcal{T}_p}(\hat{{\rm z}}_1\ch r + \hat{{\rm z}}_2^\dagger\sh r)
        + \sqrt{\mathcal{A}_p}\,\hat{{\rm v}}_1 \,, \\
    \hat{{\rm b}}_2 = \hat{{\rm a}}_2 
      = \sqrt{\eta_p}(\hat{{\rm z}}_2\ch r + \hat{{\rm z}}_1^\dagger\sh r)
        + \sqrt{1-\eta_p}\,\hat{{\rm v}}_2 \,,
  \end{gather}
\end{subequations}
where
\begin{gather}
  \mathcal{T} = 1-\mathcal{A} \,, \\
  \mathcal{T}_p = \eta_p\mathcal{T} \,, \qquad \mathcal{A}_p = 1- \eta_p\mathcal{T} \,,
\end{gather}
\begin{subequations}
  \begin{gather}
    \hat{{\rm v}}_1 = \frac{
        \sqrt{(1-\eta_p)\mathcal{T}}\,\hat{{\rm v}}_{1p} 
        + \sqrt{\mathcal{A}}\,\hat{{\rm v}}_{\rm obj}}{\sqrt{\mathcal{A}_p}} \,, \\
    \hat{{\rm v}}_2 = \hat{{\rm v}}_{2p}\, ,    
  \end{gather}
\end{subequations}
and $\hat{{\rm v}}_{\rm obj}$ is the vacuum noise introduced by the object. Note that $\hat{{\rm v}}_{1,2}$ again are two independent vacuum fields.

Finally, the DOPAs give
\begin{gather}
  \hat{{\rm c}}_1 = \hat{{\rm b}}_1\ch R + \hat{{\rm b}}_1^\dagger\sh R 
    = \sqrt{\mathcal{T}_p}(\hat{C}_1 + \hat{S}_1^\dagger)
      + \sqrt{\mathcal{A}_p}(\hat{{\rm v}}_1\ch R + \hat{{\rm v}}_1^\dagger\sh R)\,,
      \\
  \hat{{\rm c}}_2 = \hat{{\rm b}}_2\ch R + \hat{{\rm b}}_2^\dagger\sh R 
    = \sqrt{\eta_p}(\hat{C}_2 + \hat{S}_2^\dagger)
      + \sqrt{1-\eta_p}(\hat{{\rm v}}_2\ch R + \hat{{\rm v}}_2^\dagger\sh R) \,,
\end{gather}
where
\begin{subequations}
  \begin{gather}
    \hat{C}_{1,2} = \hat{{\rm z}}_{1,2}\ch r\ch R + \hat{{\rm z}}_{2,1}\sh r\sh R \,, \\
    \hat{S}_{1,2} = \hat{{\rm z}}_{1,2}\ch r\sh R + \hat{{\rm z}}_{2,1}\sh r\ch R \,. 
  \end{gather}
\end{subequations}
It follows from these equations that the mean values and the second moments of the photon numbers in the output beams are 
\begin{subequations}\label{PDC_stat_c} 
  \begin{gather}
    \mean{\hat{n}_{c1}} = \left(\mathcal{T}N + \frac{1}{2}\right)\ch2R - \frac{1}{2} \,, \\
    \mean{\hat{n}_{c2}} = \left(N + \frac{1}{2}\right)\ch2R - \frac{1}{2} \,, \\
    \mean{(\delta\hat{n}_{c1})^2} = \left(\mathcal{T}N + \frac{1}{2}\right)^2\ch4R 
      - \frac{1}{4} \,, \\
    \mean{(\delta\hat{n}_{c2})^2} = \left(N+\frac{1}{2}\right)^2\ch4R - \frac{1}{4} \,,\\ 
    \mean{\delta\hat{n}_{c1}\cdot\delta\hat{n}_{c2}} = \mathcal{T}N(N+\eta_p)\ch4R \,,
  \end{gather}
\end{subequations}
where
\begin{equation}
  N = \eta_p\sh^2r 
\end{equation} 
is the mean photon number at the object.

In the case of estimator \eqref{est_simple}, the absorption can be measured with the uncertainty [see Eqs.\,\eqref{stat_c2d}]
\begin{multline}\label{delta_A_pdc_simple} 
  (\Delta\mathcal{A})^2 
  = \frac{\mean{(\delta\hat{n}_{d1} - \delta\hat{n}_{d2})^2}}{G^2} =
  \\ \frac{1}{(G/\eta_d)^2}\left[
        \mean{(\delta\hat{n}_{c1})^2} + \mean{(\delta\hat{n}_{c2})^2}
        - 2\mean{\delta\hat{n}_{c1}\cdot\delta\hat{n}_{c2}}
        + \epsilon^2(\mean{\hat{n}_{c1}} + \mean{\hat{n}_{c2}})
      \right] .
\end{multline} 
where
\begin{equation}
  G = \partd{\mean{\hat{n}_{d1}}}{\mathcal{A}} = -\eta_dN\ch2R
\end{equation} 
is the transfer function.

In the case of estimator \eqref{est_smart}, the uncertainty is 
\begin{equation}\label{PDC_DA_raw} 
  (\Delta\mathcal{A})^2 
  = \frac{\mean{(\delta\hat{n}_{d1} - k\delta\hat{n}_{d2})^2}}{G^2} 
  = \frac{
        \mean{(\delta\hat{n}_{d1})^2} 
        - 2k\mean{\delta\hat{n}_{d1}\cdot\delta\hat{n}_{d2}}
        + k^2\mean{(\delta\hat{n}_{d2})^2}
      }{G^2} \,.
\end{equation} 
where $k$ is the factor which has to be optimized. It is easy to see that the minimum of \eqref{PDC_DA_raw} occurs at
\begin{equation}
  k = \frac{\mean{\delta\hat{n}_{d1}\cdot\delta\hat{n}_{d2}}}
    {\mean{(\delta\hat{n}_{d2})^2}}  \,,
\end{equation} 
and is [see Eqs.\,\eqref{stat_c2d}]
\begin{multline}\label{delta_A_pdc_smart} 
  (\Delta\mathcal{A})^2 = \frac{1}{G^2}\left[
      \mean{(\delta\hat{n}_{d1})^2} 
      - \frac{\mean{\delta\hat{n}_{d1}\cdot\delta\hat{n}_{d2}}^2}
          {\mean{(\delta\hat{n}_{d2})^2}}
    \right] 
  = \\ \frac{1}{(G/\eta_d)^2}\left[
      \mean{(\delta\hat{n}_{c1})^2} + \epsilon^2\mean{n_{c1}}
      - \frac{\mean{\delta\hat{n}_{c1}\cdot\delta\hat{n}_{c2}}^2}
          {\mean{(\delta\hat{n}_{c2})^2} + \epsilon^2\mean{n_{c1}}}
    \right] . 
\end{multline}

In the case of $R=0$ and (\ref{assump1}, \ref{big_N}), Eqs.\,(\ref{delta_A_pdc_simple}, \ref{delta_A_pdc_smart}) simplify, respectively, to Eqs.\,(\ref{Delta2Apdc1_0}, \ref{Delta2Apdc2_0}).

In the case of $e^R\gg1$ and \eqref{big_N}), Eqs.\,(\ref{delta_A_pdc_simple}, \ref{delta_A_pdc_smart}) give, respectively,
\begin{gather}
  (\Delta\mathcal{A})^2 = 2\biggl\{
      \mathcal{A}^2 
      + \frac{1}{N}\left[
            \mathcal{A} + 2(1-\eta_p)\mathcal{T} + \epsilon^2(1+\mathcal{T})e^{-2R}
          \right]  
    \biggr\} + \frac{1}{N^2} \,, \\
  (\Delta\mathcal{A})^2 = \frac{2\mathcal{T}}{N}
    \left[\mathcal{A} + 2(1-\eta_p)\mathcal{T} + \epsilon^2(1+\mathcal{T})e^{-2R}\right] 
    + \frac{1}{N^2}
        \left[\frac{1+\mathcal{T}^2}{2} + 2\mathcal{T}(1-\mathcal{T}\eta_p^2)\right] .
\end{gather}
If, in addition, assumptions \eqref{assump1} are fulfilled, then these equations simplify, respectively, to Eqs.\,(\ref{Delta2Apdc1_R}, \ref{Delta2Apdc2_R}). 

\section{Squeezed coherent state}\label{app:coh_sqz}

\begin{figure}
  \includegraphics[width=0.85\textwidth]{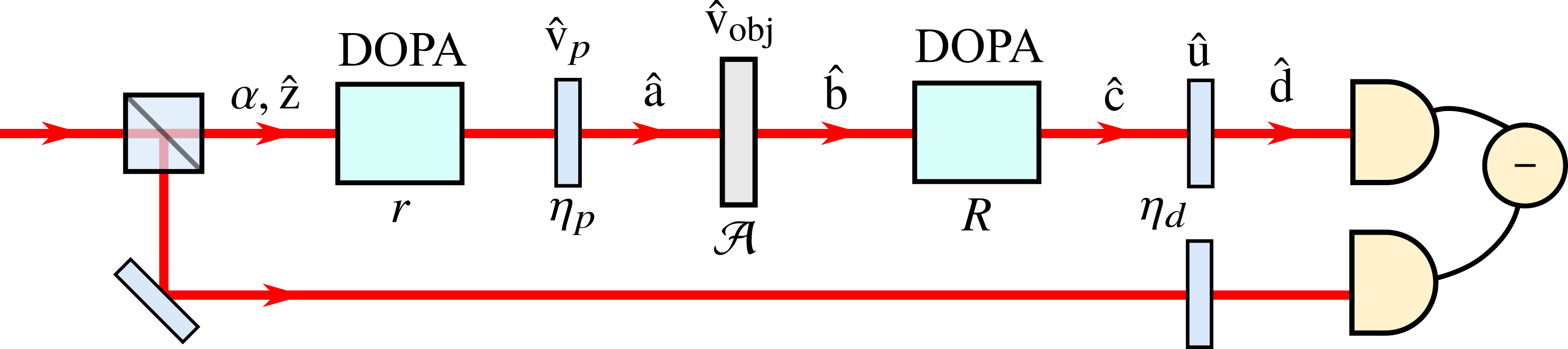}
  \caption{Measurement of absorption using a squeezed coherent state and amplification before detection. Since the reference beam is considered to have a very large amplitude in order to suppress the shot-noise in the corresponding channel, one can take into account only the noise of the signal channel. Here $\alpha$ is rescaled, see Eq.\,(\ref{eq:SqzCohAlpha}).}
  \label{fig:scheme_SQZ_fields} 
\end{figure} 

With an account for the preparation imperfection, the annihilation operator for the field incident at the object is 
\begin{equation}\label{z2a_sqz} 
  \hat{{\rm a}} = \alpha + \sqrt{\eta_p}(\hat{{\rm z}}\ch r + \hat{{\rm z}}^\dagger\sh r)
    + \sqrt{1-\eta_p}\,\hat{{\rm v}}_p\,,
\end{equation} 
wnere $\hat{{\rm z}}$, $\hat{{\rm v}}_p$ are vacuum fields  and 
\begin{equation} \label{eq:SqzCohAlpha}
  \alpha = \sqrt{N - \eta_p\sh^2r} 
\end{equation} 
(we assume that $\alpha$ is real). The object modifies this field as follows:
\begin{equation}
  \hat{{\rm b}}
  = \sqrt{\mathcal{T}}\hat{{\rm a}} + \sqrt{\mathcal{A}}\hat{{\rm v}}_{\rm obj}
  = \sqrt{\mathcal{T}}\alpha 
    + \sqrt{\mathcal{T}_p}(\hat{{\rm z}}\ch r + \hat{{\rm z}}^\dagger\sh r)
    + \sqrt{\mathcal{A}_p}\,\hat{{\rm v}} \,,
\end{equation} 
where 
\begin{gather}
  \hat{{\rm v}} = \frac{
      \sqrt{(1-\eta_p)\mathcal{T}}\,\hat{{\rm v}}_p 
      + \sqrt{\mathcal{A}}\,\hat{{\rm v}}_{\rm obj}
    }{\sqrt{\mathcal{A}_p}} 
\end{gather} 
and $\hat{{\rm v}}_{\rm obj}$ is the vacuum noise introduced by the object. Note that $\hat{{\rm v}}$ again is a vacuum noise. 

Finally, the DOPA gives
\begin{multline}\label{SQZ_c} 
  \hat{{\rm c}} = \hat{{\rm b}}\ch R + \hat{{\rm b}}^\dagger\sh R \\
  = \sqrt{\mathcal{T}}\alpha e^R 
    + \sqrt{\mathcal{T}_p}
        \bigl[\hat{{\rm z}}\ch(r+R) + \hat{{\rm z}}^\dagger\sh(r+R)\bigr] 
    + \sqrt{\mathcal{A}_p}(\hat{{\rm v}}\ch R + \hat{{\rm v}}^\dagger\sh R) \,.
\end{multline} 
It follows from \eqref{SQZ_c} that the mean value and the variance of the photon number in the output beam are 
\begin{align} 
	\mean{\hat{n}_{c}} &
      = \mathcal{T}\alpha^2e^{2R} + \mathcal{T}_p\sh^2(r+R) + \mathcal{A}_p\sh^2R \,, \\
      \begin{split}
  \mean{(\delta\hat{n}_c)^2} = & \mathcal{T}\mathcal{T}_p\alpha^2e^{2r+4R} +
     \frac{\mathcal{T}_p^2}{2}\sh^22(r+R) 
      + \\ & \mathcal{T}\mathcal{A}_p\alpha^2e^{4R} + \mathcal{T}_p\mathcal{A}_p\sh^2(r+2R) 
      + \frac{\mathcal{A}_p^2}{2}\sh^22R \,.
      \end{split}
\end{align}

With an account for Eqs.\,\eqref{stat_c2d}, the absorption measurement error is 
\begin{equation}\label{delta_A_coh_sqz} 
  (\Delta\mathcal{A})^2 
  = \frac{\mean{(\delta\hat{n}_d)^2}}{G^2} 
  = \frac{\mean{(\delta\hat{n}_c)^2} + \epsilon^2\mean{\hat{n}_c}}{(G/\eta_d)^2} \,,
\end{equation} 
where
\begin{equation}
  G = \partd{\mean{\hat{n}_d}}{\mathcal{A}} 
  = \eta_d\bigl[-\alpha^2e^{2R} - \eta_p\sh^2(r+R) + \eta_p\sh^2R\bigr]
\end{equation} 
is the transfer function.


\end{document}